\documentclass[twocolumn,english,superscriptaddress]{revtex4-1}
\usepackage[T1]{fontenc}
\usepackage{amsfonts}
\usepackage{amsmath}
\usepackage[latin9]{inputenc}
\usepackage{textcomp}
\usepackage{amstext}
\usepackage{subscript}
\usepackage{graphics}
\usepackage{graphicx}
\usepackage{float}

\makeatletter

\newcommand{\lyxmathsym}[1]{\ifmmode\begingroup\def\b@ld{bold}
  \text{\ifx\math@version\b@ld\bfseries\fi#1}\endgroup\else#1\fi}

\usepackage{hyperref}
\usepackage{upgreek,textgreek}
\hypersetup{colorlinks=true,linkcolor=blue,citecolor=blue,urlcolor=blue}
\makeatother

\usepackage{babel}
\begin{document}
\title{Magneto-Seebeck effect and ambipolar Nernst effect in CsV\textsubscript{$3$}Sb\textsubscript{5}
superconductor}
\author{Yuhan Gan}
\thanks {These authors contributed equally to this work.}
\affiliation{Low Temperature Physics Lab, College of Physics \& Center of Quantum
Materials and Devices, Chongqing University, Chongqing 401331, China}
\author{Wei Xia \textcolor{blue}{$^*$}}

\affiliation{School of Physical Science and Technology, ShanghaiTech University, Shanghai 201210, China}
\affiliation{
 ShanghaiTech Laboratory for Topological Physics, Shanghai 201210, China}
\author{Long Zhang\textcolor{blue}{$^*$}}
\affiliation{Low Temperature Physics Lab, College of Physics \& Center of Quantum
Materials and Devices, Chongqing University, Chongqing 401331, China}
\author{Kunya Yang}
\affiliation{Low Temperature Physics Lab, College of Physics \& Center of Quantum
Materials and Devices, Chongqing University, Chongqing 401331, China}
\author{Xinrun Mi}
\affiliation{Low Temperature Physics Lab, College of Physics \& Center of Quantum
Materials and Devices, Chongqing University, Chongqing 401331, China}
\author{Aifeng Wang}
\affiliation{Low Temperature Physics Lab, College of Physics \& Center of Quantum
Materials and Devices, Chongqing University, Chongqing 401331, China}
\author{Yisheng Chai}
\affiliation{Low Temperature Physics Lab, College of Physics \& Center of Quantum
Materials and Devices, Chongqing University, Chongqing 401331, China}
\author{Yanfeng Guo}
\email{guoyf@shanghaitech.edu.cn}
\affiliation{School of Physical Science and Technology, ShanghaiTech University, Shanghai 201210, China}
\author{Xiaoyuan Zhou}
\email{xiaoyuan2013@cqu.edu.cn}

\affiliation{Low Temperature Physics Lab, College of Physics \& Center of Quantum
Materials and Devices, Chongqing University, Chongqing 401331, China}
\author{Mingquan He}
\email{mingquan.he@cqu.edu.cn}

\affiliation{Low Temperature Physics Lab, College of Physics \& Center of Quantum
Materials and Devices, Chongqing University, Chongqing 401331, China}
\date{today}

\begin{abstract}
We present a study of Seebeck and Nernst effect in combination with magnetoresistance and Hall measurements of the Kagome superconductor CsV\textsubscript{3}Sb\textsubscript{5}. Sizable magneto-Seebeck signal appears once the charge density wave (CDW) order sets in below $T_{CDW}$=94 K.  The Nernst signal peaks at a  lower temperature $T^*\sim$35 K, crossing which the Hall coefficient switches sign, which we attribute to the ambipolar transport of compensated bands due to the multi-band nature of CsV\textsubscript{3}Sb\textsubscript{5}. Sublinear Nernst signal as a function of magnetic field, together with large anomalous Nernst effect (ANE) also emerge inside the CDW phase, despite the absence of long-range magnetic order. These findings suggest that, the transport properties are dominated by small pockets with multi-band profile,  and that the unusual band topology also plays an import role in the CDW state of  CsV\textsubscript{3}Sb\textsubscript{5}.

\end{abstract}

\maketitle

\section{Introduction}
Intertwined orders,  such as magnetic, charge, nematic and superconducting orders often lead to rich phases in quantum materials \cite{Fernandes2019}. Involvement of non-trivial band topology can further enrich the physics as evidenced by the advent of magnetic topological materials \cite{Tokura2019}, topological superconductors \cite{Sato2017}, etc. The recent discovered superconducting Kagome family \textit{A}V\textsubscript{3}Sb\textsubscript{5} (\textit{A}=K, Rb, Cs) provides another prime example to investigate the interplay between intertwined orders and non-trivial band topology \cite{Ortiz2019,Ortiz2020,Ortiz2021,Yin2021}. Both charge density wave (CDW) \cite{Jiang2021a,Liang2021CDW,Zhao2021a,Chen2021rotonpair,Xu2021,Li2021a,Shumiya2021,Wang2021CDW}, superconductivity \cite{Ortiz2019,Ortiz2020,Ortiz2021,Yin2021}, nematicity \cite{Xiang2021} and Dirac-like bands \cite{Ortiz2019,Ortiz2020,Li2021,Nakayama2021,Wang2021Cs,Liu2021,Ortiz2021Cs,Hu2021} have been identified in these compounds. Theoretically,  chiral flux CDW phase \cite{Feng2021,Denner2021} and unconventional superconducting pairing \cite{Ko2009,Yu2012,Kisel2013} could be expected due to the intimate coupling between charge order, superconductivity and topological band structure.   Indeed, experimental signatures of chiral charge order \cite{Jiang2021a,Zhao2021a,Wang2021CDW,Yu2021b,Mielke2021}, unconventional superconductivity \cite{Wang2020,Zhao2021,Xiang2021}, roton pair density wave \cite{Chen2021rotonpair} and Majorana excitation \cite{Liang2021CDW} were reported, which have triggered intensive research effort to unveil the nature of CDW and superconductivity, and to search for non-trivial states in this Kagome system \cite{Jiang2021}.  

The \textit{A}V\textsubscript{3}Sb\textsubscript{5} compounds crystallize into a layered structure (space group: $P6/mmm$) with V-Sb layers sandwiched by \textit{A} atoms \cite{Ortiz2019}.   At room temperature, a perfect Kagome lattice is formed in the V-Sb layer by V atoms, which gives rise to multiple Dirac points near the Fermi level ($E_F$) and the $\mathbb{Z}_2$ topology in the electronic band structure \cite{Ortiz2020}.   Both density functional (DFT) calculations and angle-resolved photoemission spectroscopy (ARPES) experiments have suggested a mulitband nature of the band structure \cite{Ortiz2019,Ortiz2020,Tan2021,Ortiz2021Cs,Li2021,Nakayama2021,Wang2021Cs,Liu2021,Ortiz2021Cs,Hu2021}.  An electron pocket is located at Brillouin zone (BZ) center ($\varGamma$ point),  involving the $p_z$ orbital of in-plane Sb atoms. The $d$ orbitals of V atoms intersect with the Fermi level at the BZ boundaries, forming both electron ($k_z=\pi$, $L$ point) and hole ($k_z=0$) bands at the $M$ point \cite{Li2021}.  Additionally, two van Hove (VH) singularity points (saddle points) pop up near $E_F$ around the $M$ point. The CDW phase emerges in all three systems below a certain transition temperature  $T_{CDW}$ ($T_{CDW}\approx$ 78, 103, 94 K for K, Rb, Cs based compounds, respectively) \cite{Ortiz2019,Ortiz2020,Ortiz2021,Yin2021}. Consequently,  substantial Fermi surface reconstruction occurs mainly on the pockets sitting at the $M$ point, which is accompanied by moderate lattice distortions \cite{Ortiz2021Cs}. The CDW modulation appears to be three-dimensional with a in-plane 2$\times$2 superstructure, while the out-of-plane periodicity is under debate \cite{Jiang2021}.   The origin of the CDW instability is also unclear. A few scenarios, including Fermi surface nesting among the saddle points, quasi-nesting between electron and hole pockets  at $L$ and $M$ points,  have been proposed \cite{Tan2021,Ortiz2021Cs,Li2021}.  

Intriguingly, a giant anomalous Hall effect (AHE) has been reported in the CDW phase of K\textsubscript{1-x}V\textsubscript{3}Sb\textsubscript{5} \cite{Yang2020} and CsV\textsubscript{3}Sb\textsubscript{5}   \cite{Yu2021}, despite the absence of long-range magnetic order or local moments \cite{Kenney2021}. The origin of the observed giant AHE  is still a mystery, which was initially attributed to extrinsic skew scattering of spin clusters \cite{Yang2020}.  Another more exotic possibility could be the emergent effect of time reversal symmetry breaking, which is supported by the observation of chrial charge order in scanning tunneling microscopy (STM) \cite{Jiang2021a,Zhao2021a,Wang2021CDW}  and muon spin relaxation/rotation ($\mu$SR) \cite{Yu2021b,Mielke2021} experiments. However, the connection between the observed AHE and chiral charge order is far from evident.  Notably, the appearance of AHE was found to be concomitant with the CDW transition \cite{Yu2021}, suggesting an intrinsic role played by the reconstruction of Fermi surface topology at $T_{CDW}$.  The transport study is , however, hitherto limited to the resistivity channel. The thermopower and Nernst effect can probe the entropy flow carried by quasi-particles and are very sensitive to unusual band topology near the Fermi level \cite{Xiao2006prl,XiaoRevMod}, which have been widely applied to the study of intertwined phases of unconventional superconductors \cite{Behnia2016} and topological materials \cite{Fuchengguan}.  Further investigation using Seebeck and Nernst study can extract more information concerning the role played by band topology. 

In this report, we performed magneto-Seebeck and Nernst measurements on single crystalline CsV\textsubscript{3}Sb\textsubscript{5} superconductors. The thermopower is dominated by electron-like carriers but with distinct energy scales above and below the CDW transition. Above $T_{CDW}$ =94 K, the transport is likely dominated by the electron band at the $\Gamma$ point, while the small pockets at the $M$ point play a major role in the CDW phase.  Appreciable magneto-Seebeck effect develops concomitantly with the entrance of CDW order, which also points to a small energy scale that can be easily tuned by magnetic field. Moreover, sizable Nernst signal is observed in the CDW state, which peaks at the temperature ($T^*\sim$35 K) where the Hall coefficient changes sign. We argue that this is originated from the ambipolar transport of compensated  electron-like and hole-like bands at $L$ and $M$ points, similar to that in 2$H$-NbSe\textsubscript{2} \cite{Bel2003}. Desipte the absence of long-range magnetic order, clear signatures of anomalous Nernst effect are also found in the CDW phase, which have comparable values to those in magnetic topological materials \cite{YangHaiyang2020,Guin2019}. Our findings suggest that, the particle-hole scattering among $L$ and $M$ points, and the band topology play important roles in the transport properties of the CDW phase in CsV\textsubscript{3}Sb\textsubscript{5} .

\section{Methods}
Single crystalline CsV\textsubscript{3}Sb\textsubscript{5} samples were grown by the self-flux method as described elsewhere \cite{Zhao2021}. Magnetization measurements were carried out using a
Quantum Design Magnetic Properties Measurement System (MPMS3). The Seebeck, Nernst, magnetoresistance and Hall measurements were performed in a 14-T Oxford cryostat. Longitudinal resistivity and magnetoresistance were measured using the standard four-probe method. Hall measurements were performed using a five-terminal configuration.  One heater, two-thermometer geometry was used to capture the thermopower and Nernst effect. Two Keithley 2182A nanovoltmeters were used to record the Seebeck and Nernst signals at the same time.

\section{Results And Discussions}
\begin{figure*}[t]
\centering
\includegraphics[scale=0.7]{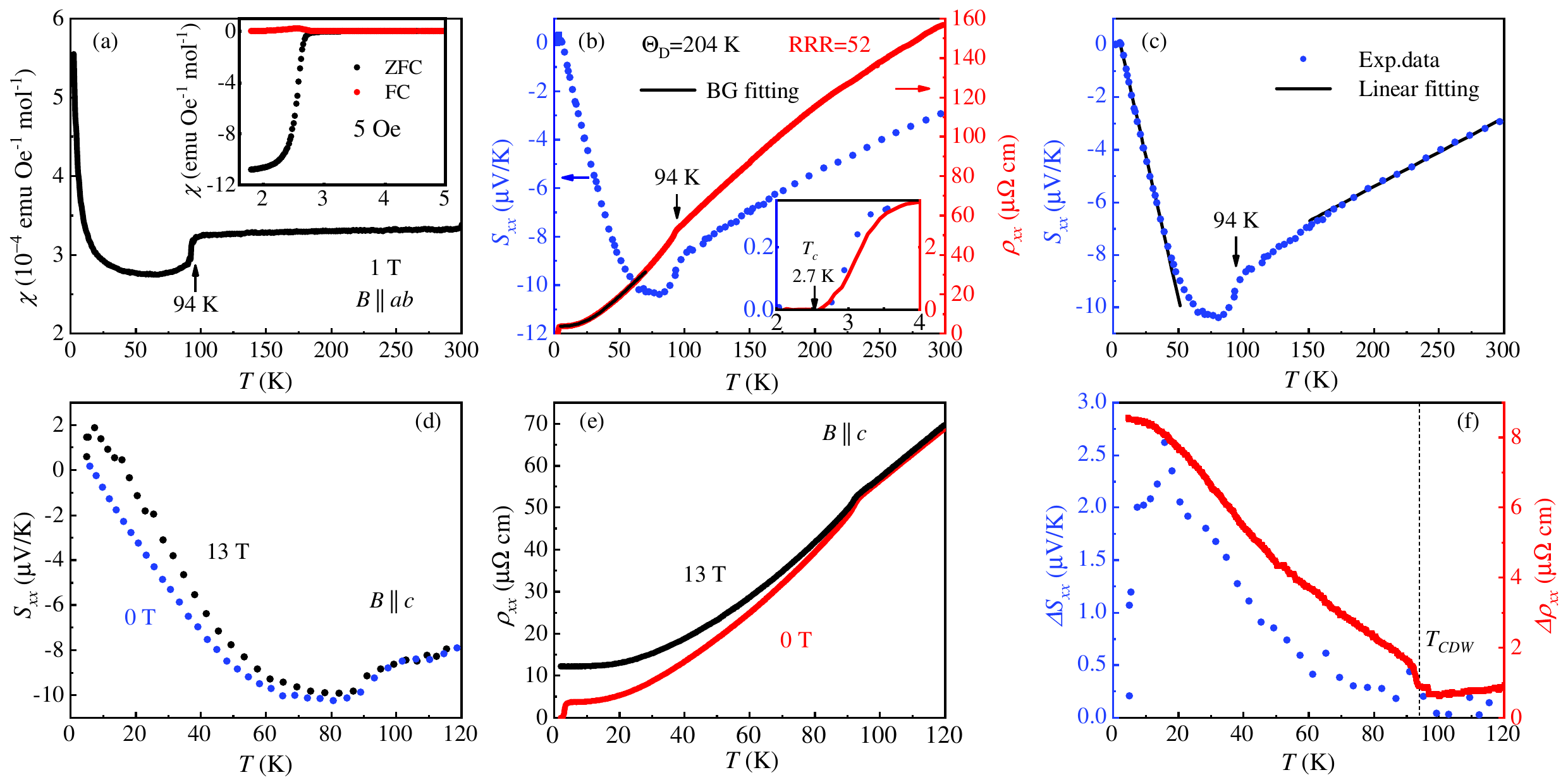}
\caption{(a)(b)Temperature dependence of the magnetization, in-plane Seebeck coefficient and resistivity of CsV\textsubscript{3}Sb\textsubscript{5}. Both the CDW ($T_{CDW}$=94 K) and superconductivity [$T_c$=2.7 K, see the insets in (a)(b)] transitions are nicely resolved. The black solid line in (b) is a Bloch-Gr\"uneisen (BG) fitting of the normal state resistivity from 5 to 70 K. The Debye temperature is estimated to be $\theta_D$=204 K. (c) Linear scaling of $S_{xx}$ well above (150-300 K) and below (10-40 K) the CDW transition. (d)(e) Seebeck effect and resistivity measured at zero magnetic field and $B$=13 T with field applied along the $c$-axis. (f) The magneto-Seebeck effect $\Delta S_{xx}$=$S_{xx}$(13 T)-$S_{xx}$(0 T) and magnetoresistance $\Delta \rho_{xx}$=$\rho_{xx}$(13 T)-$\rho_{xx}$(0 T) evaluated using the data shown in (d) and (e). }
\label{fig1}
\end{figure*}
In Figure \ref{fig1}, we present the magnetization, in-plane Seebeck signal $S_{xx}$, together with longitudinal resistivity $\rho_{xx}$ of a CsV\textsubscript{3}Sb\textsubscript{5} single crystal. For magnetization experiments, the magnetic field was applied parallel to the $ab$ plane ($B\parallel ab$). Both zero field cooled (ZFC) and field cooled (FC) measurements were performed in the vicinity of superconducting transition [see the inset of  Fig. \ref{fig1}(a)]. For electrical transport and thermopower measurements, the electrical current and thermal gradient were applied within the $ab$ plane, while the magnetic field was directed along the $c$-axis ($B\parallel c$). As shown in Figs.\ref{fig1}(a) and (b), the CDW transition is manifested by a kink at 94 K in resistivity, which is much more pronounced in magnetization and thermopower. Superconductivity occurs below $T_c$=2.7 K, characterized by Meissner effect, zero resistance and zero thermopower [see the insets of Figs. \ref{fig1}(a) and (b)]. The residual resistivity ratio (RRR), defined as $\rho_{xx}$(300 K)/$\rho_{xx}$(5 K), equals to 52, implying the high quality of our crystal. Both the CDW and superconducting transition temperatures math well with those reported earlier \cite{Ortiz2020,Yu2021}. The Seebeck signal retains a negative sign all the way down to about 6 K, below which  small positive values are found until $T_c$ is reached. This implies that the thermopower is dominated by electron-like carriers. Followed by the sudden drop at $T_{CDW}$, the thermopower reaches a broad maximum around 70 K, which is likely a consequence of phonon drag contribution. The phonon drag peak typically appears around $\theta_D$/5 ($\theta_D$: Debye temperature) in metals, such as Cu, Ag, Au, Al \cite{Barnard1972}. Here, we estimate the Debye temperature by fitting the normal state resistivity according to the Bloch-Gr\"uneisen (BG) model, assuming  dominating phonon-scattering contribution \cite{Gruneisen:1933aa}:
\begin{equation}
\begin{aligned}
\rho(T)&=  \rho_{0}+\rho_{ph}(T)\\& =\rho_{0}+\frac{4A}{\theta_{D}}\left(\frac{T}{\theta_{D}}\right)^{n}\int_{0}^{\theta_{D}/T}\frac{z^{n}dz}{(e^{z}-1)(1-e^{-z})}-kT^{3}.
\label{eq1}
\end{aligned}
\end{equation}As shown in Fig. \ref{fig1}(b), the normal state resistivity can be well described by the Bloch-Gr\"uneisen approach (the solid back line) up to 70 K, which  yields a residual resistivity $\rho_0$=3.8 $\mu\Omega$ cm, Debye temperature $\theta_D$=204 K, electron-phonon coupling constant $A$=11 m$\Omega$ cm K. Typical values of the exponent $n$ are 2, 3 or 5 \cite{Bid2006BG}, and using $n=3$ gives the best fitting in our case.  The cubic term $kT^3$ turns out to be negligible in our analysis. Using the same treatment for  CsV\textsubscript{3}Sb\textsubscript{5},  X. Chen \textit{et al}. \cite{Chen2021a} obtained similar magnitude of Debye temperature, which is slightly higher than that in   KV\textsubscript{3}Sb\textsubscript{5} ($\theta_D$=141 K \cite{Ortiz2021K}) and RbV\textsubscript{3}Sb\textsubscript{5} ($\theta_D$=169 K, \cite{Yin2021}). 

As clearly seen in Fig. \ref{fig1}(c), the Seebeck signal scales linearly with temperature well above and below the CDW transition,  where the phonon assisted transport fades away. From the linear scaling, one can extract the characteristic energy scale involved in the entropy flow. The Seebeck effect dominated by diffusive transport can be modeled by the Mott relation \cite{Mott1936}:
\begin{equation}
\begin{aligned}
S_{d}=\left.\frac{\pi^{2}}{3}\frac{k_{B}^{2}T}{e}\left(\frac{\partial\,\mathrm{ln}\sigma(E)}{\partial E}\right)\right|_{E_{F}},
\label{eq2}
\end{aligned}
\end{equation}
where $e$ is the electronic charge of carrier ($e$<0 for electrons and $e$>0 for holes), $k_B$ is the Boltzman constant, $T$ is the Kelvin temperature, $E$ is the energy of charge carriers, and $\sigma(E)$ is the energy dependent electrical conductivity. The Fermi energy $E_F$ is kept positive by counting upward (downward) for electron (hole) band. In the simple case of isotropic scattering by impurities, Eq. \ref{eq2}  can be simplified to:
\begin{equation}
\begin{aligned}
S_{d}=C\frac{\pi^{2}}{e}\frac{k_{B}^{2}}{E_{F}}T,
\end{aligned}
\end{equation}which is a linear function of the temperature $T$. The constant $C$ is 1/3 and 1/6 for three-dimensional and two-dimensional Fermi surfaces, respectively \cite{abrikosov1988fundamentals}. Here, we take $C$=1/3 considering the quasi-2D nature of the band structure \cite{Tan2021,Ortiz2021Cs}. At high-temperature from 150 K to 300 K, the linear scaling yields $E_F$=440 meV, which is close to the energy scale of the electron band centered at the $\Gamma$ point \cite{Ortiz2020,Ortiz2021Cs,Tan2021,Li2021}. Deep inside the CDW phase from 10 K to 40 K, on the other hand, gives a much smaller value of $E_F$=60 meV, which agrees well with the size of the electron pocket at the $L$ point ($\sim$ 50 meV) as resolved by ARPES study \cite{Yin2021}.  These findings suggest that the dominant roles played by carriers from different bands are switched over by substantial Fermi surface reconstruction occurring at the CDW transition \cite{Ortiz2021Cs}. The transport properties in the CDW phase are mainly controlled by the small pockets around the zone boundaries.

The different carrier dynamics above and below $T_{CDW}$ can be further evidenced by different response to magnetic field as shown in Figs. \ref{fig1}(d)-(f). Sizable magneto-Seebeck and magnetoresistance only develop inside the CDW phase, which is seen more clearly by taking the difference between data recorded at zero field and 13 T [see Fig. \ref{fig1}(f)]. The magneto-Seebeck response $\Delta S_{xx}$=$S_{xx}$(13 T)-$S_{xx}$(0 T) is practically zero above  $T_{CDW}$ within our measurement uncertainties.  The magnitude of $\Delta S_{xx}$ grows gradually inside the CDW, reaching a maximum around 20 K, which drops rapidly to zero approaching $T_c$. The magnetoresistance $\Delta \rho_{xx}$=$\rho_{xx}$(13 T)-$\rho_{xx}$(0 T) remains finite but nearly temperature independent above $T_{CDW}$. It rises gradually by entering the CDW phase and eventually saturates below 10 K.  The appearance of magnetic response inside the CDW phase again suggests a small energy scale which is tunable by magnetic fields. Similar effects have also been found in the CDW state of P\textsubscript{4}W\textsubscript{12}O\textsubscript{44} possessing small Fermi pockets \cite{Kolincio}.

\begin{figure*}[t]
\centering
\includegraphics[scale=0.6]{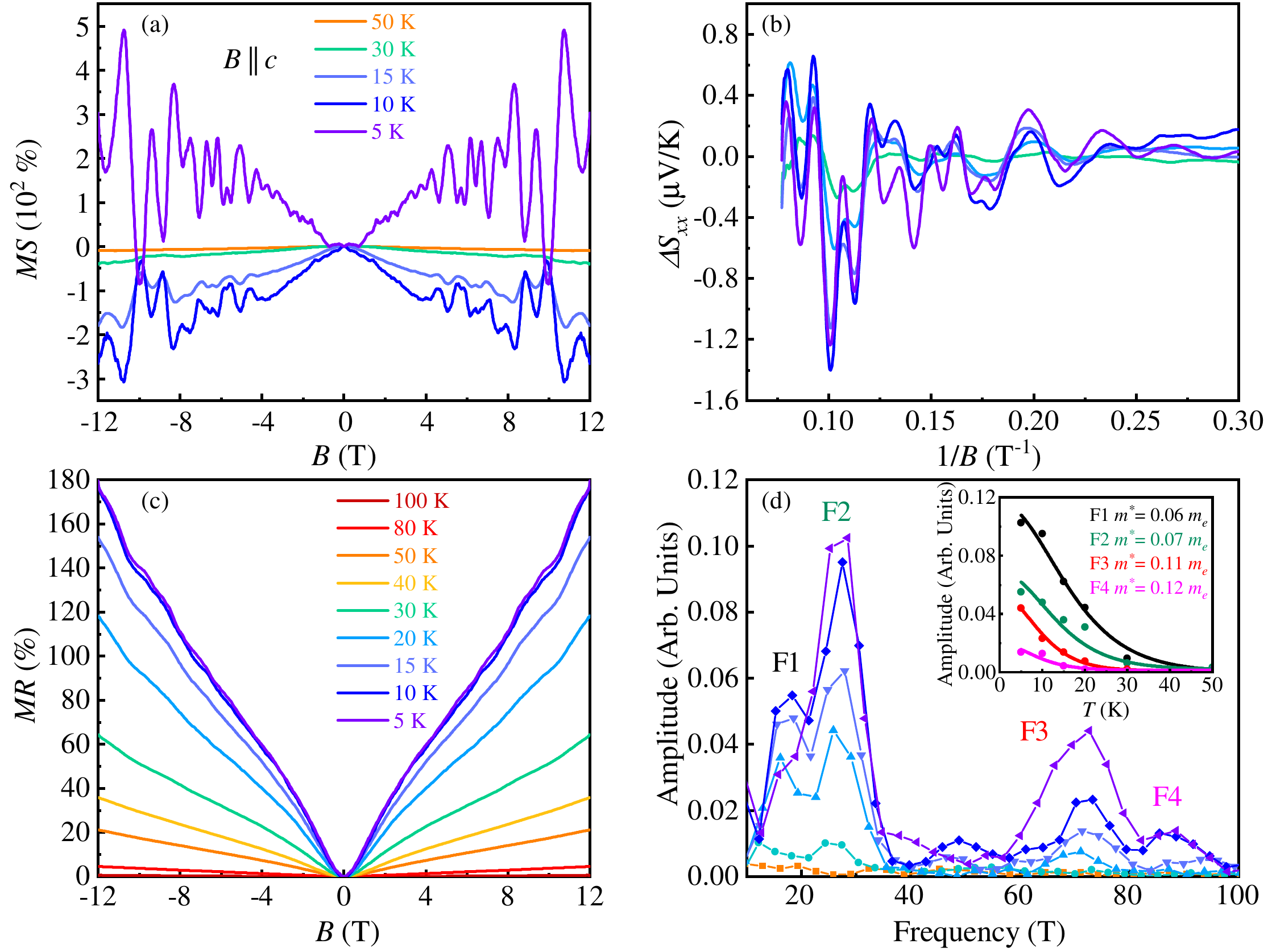}
\caption{Magneto-Seebeck $MS=[S_{xx}(B)-S_{xx}(0)]/S_{xx}(0)$ (a) and magnetoresistance $MR=[\rho_{xx}(B)-\rho_{xx}(0)]/\rho_{xx}(0)$ (c) measured at various temperatures with $B\parallel c$. (b) The quantum oscillations $\Delta S_{xx}$ plotted as a function of $1/B$. The data is obtained by subtracting a polynomial background between 2 to 12 T from the $MS$ results shown in (a). (d) Extracted frequency from the FFT analysis of the oscillation data in (b). Four frequencies with $F_1$=18, $F_2$ =28, $F_3$= 72 and $F_4$=88 T are found. The inset shows the corresponding  effective mass  $m^*$ estimated using the Lifshitz-Kosevich approach. }
\label{fig2}
\end{figure*}

Figure \ref{fig2} presents more details on the magneto-Seebeck $MS=[S_{xx}(B)-S_{xx}(0)]/S_{xx}(0)$ and magnetoresistance $MR=[\rho_{xx}(B)-\rho_{xx}(0)]/\rho_{xx}(0)$ measured at various temperatures using $B\parallel c $. Clear quantum oscillations (QOs) are seen both in Seebeck and resistivity channels below 30 K, as shown in Figs. \ref{fig2}(a) and (c). Compared to resistivity, the Seebeck signal appears to be more sensitive to the Fermi surface topology, giving much more pronounced oscillations in the raw data. The oscillatory component of $MS$ as a function of $1/B$ is displayed in Fig. \ref{fig2}(b), by subtracting a polynomial background from 2 to 12 T.  The fast Fourier transform (FFT) analysis of the oscillations is presented in Fig. \ref{fig2}(d).  Four frequencies can be identified with $F_1$=18, $F_2$ =28, $F_3$= 72 and $F_4$=88 T, which are consistent with earlier reports extracted from QOs in magnetoresistance \cite{Yu2021,Ortiz2021Cs}.  We note that, much higher frequency up to $\sim$ 2000 T was observed in Shubnikov-de Haas (SdH) oscillations \cite{Ortiz2021Cs}, which is not evident in our data. Using the Onsager relation $F=(\hbar/2\pi e)A_F$, we can estimate the extremal area $A_F$ of the Fermi surfaces with the frequency $F$ obtained in Fig. \ref{fig2}(d). The obtained values are very small, which are 0.0017 {\AA}\textsuperscript{-2}, 0.0027  {\AA}\textsuperscript{-2}, 0.0069 {\AA}\textsuperscript{-2}, 0.0084 {\AA}\textsuperscript{-2},  for $F_1$, $F_2$, $F_3$ and $F_4$ respectively. These observations are in line with the small Fermi energy of 60 meV obtained above. The effective mass ($m^*$) can be estimated by fitting the temperature dependence of the FFT peak amplitude $a_i(T)$ for each frequency $F_i$($i=1-4$),  using the Lifshitz-Kosevich (LK) approximation $a_{i}(T)\approx X/\left(\sinh X\right)$ with $X=14.69m^*T/B$. The magnetic field $B$ (=7.25 T in our case) is typically taken as the average value of the field range used for FFT analysis. As shown in the inset of Fig. \ref{fig2}(d), the obtained values of $m^*$ for all frequencies are very small, suggesting contribution from the Dirac bands in the vicinity of the $M$ point \cite{Ortiz2021Cs}.   

\begin{figure*}[t]
\centering
\includegraphics[scale=0.7]{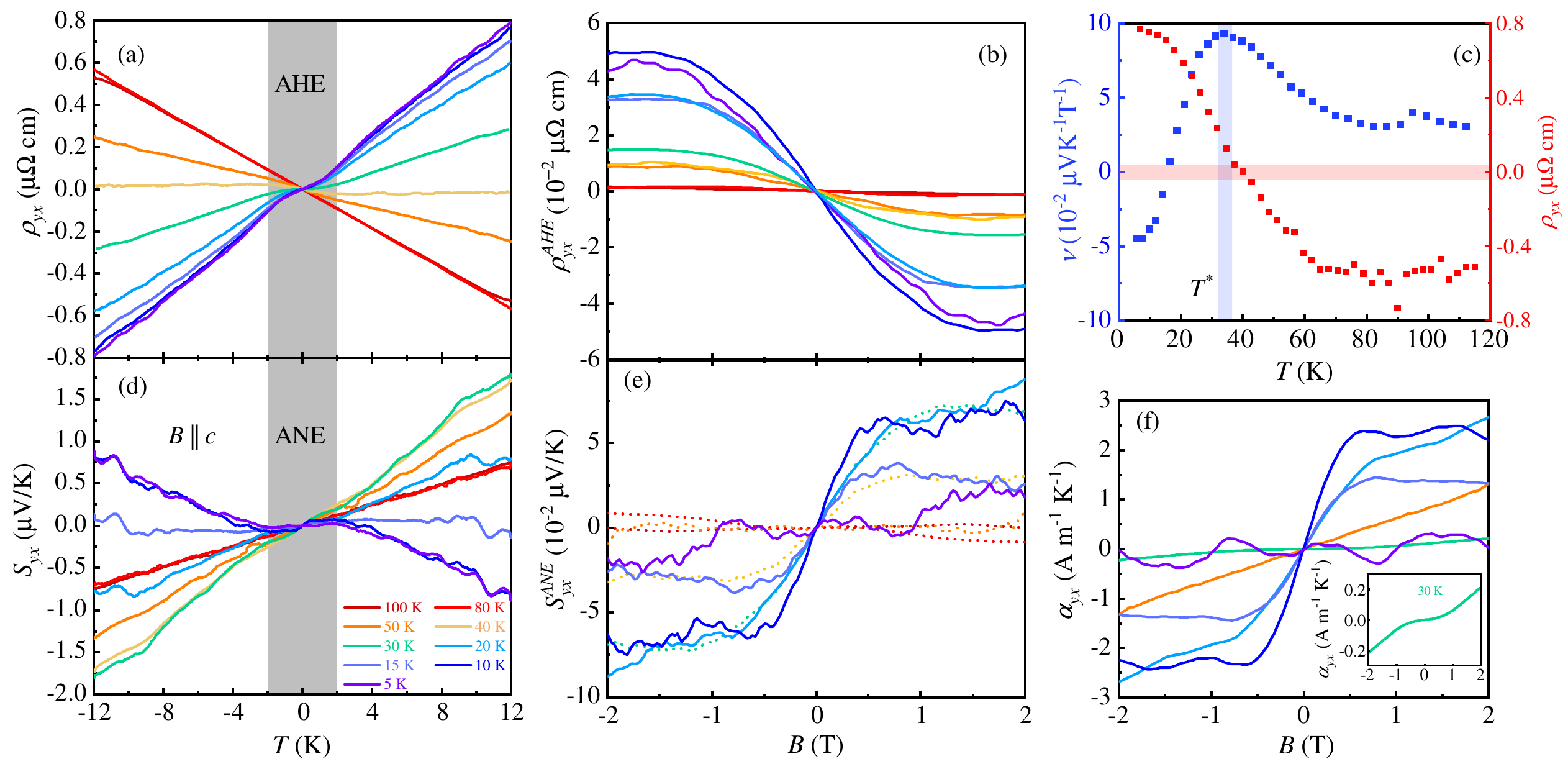}
\caption{The Hall effect (a) and Nernst effect (d) recorded at different temperatures using $B\parallel c$. The grey shaded area marks the anomalous part. (b) Enlarged view of the anomalous Hall resistivity $\rho_{yx}^{AHE}$, which is obtained by subtracting a local linear background between 1 and 2 T at each temperature. (c) Comparison of the Nernst coefficient $\nu=S_{yx}/B$ and Hall resistivity measured at $B$=12 T. The temperature $T^*\sim 35 $ K, at which $\nu$ reaches maximum and $\rho_{yx}$ switches sign, is marked out by blue and red shaded stripes. (e) Roomed in view of the anomalous Nernst signal $S_{yx}^{ANE}$. The solid lines are raw data, and the dashed lines are obtained after subtracting a local background for each curve. (f) The off-diagonal component $\alpha_{yx}$ of the thermoelectric conductivity tensor. Solid Lines are raw data. Clear signatures of ANE are seen both in  $S_{yx}^{ANE}$ and $\alpha_{yx}$ below 40 K.}
\label{fig3}
\end{figure*}

Now, we turn our attention to the transverse transport and the results are displayed in Fig. \ref{fig3}. The Hall resistivity ($\rho_{yx}$) is dominated by electron-like carriers above 80 K, resulting in a linear response to magnetic field with a negative slope [see Fig. \ref{fig3}(a)].  Sublinear effect develops at lower temperatures in the vicinity of zero field [grey shaded region in Fig. \ref{fig3}(a)], which eventually turns into an 'S'-like profile below 40 K. Note that, the Hall resistivity switches its sign to positive below 40 K, which is likely the consequence of compensated electron and hole bands as we will discuss later. Similar sign change at $\sim$35 K has also been reported earlier \cite{Yu2021}, and also appears in  K\textsubscript{1-x}V\textsubscript{3}Sb\textsubscript{5} \cite{Yang2020} and RbV\textsubscript{3}Sb\textsubscript{5} \cite{Yin2021}. The sublinear region denoted as $\rho_{yx}^{AHE}$ is amplified in Fig. \ref{fig3}(b), after subtracting a local linear background. Signatures of anomalous Hall effect appear gradually inside the CDW sate, agreeing well with earlier report \cite{Yu2021}. 

Similarly as shown in Fig. \ref{fig3}(d), the Nernst signal, $S_{yx}=E_{y}/\nabla_{x}T$, scales linearly with $B$ above 80 K and becomes sublinear at low-temperature. Notably, instead of switching sign in Hall resistivity, the Nernst signal reaches maximum around 30-40 K. In Fig. \ref{fig3}(c), we directly compare the Hall resistivity and the Nernst coefficient ($\nu=S_{yx}/B_z$) measured at $B_z=12 $ T. The Nernst coefficient,  grows gradually upon cooling and peaks at $T^*\sim$35 K. Further cooling causes rapid drop in $\nu$, which changes sign below 20 K. The Hall resistivity $\rho_{yx}$, on the other hand, switches sign around $T^*$. Similar effects, i.e., the Nernst coefficient peaks at the temperature where the Hall coefficient switches sign, has been observed in the CDW phase of 2$H$-NbSe\textsubscript{2} superconductor, which was attributed to the ambipolar transport of compensated electron and hole bands \cite{Bel2003}. Surprisingly, the maximum magnitude of $\nu(T^*)$=0.09 $\mu$V K\textsuperscript{-1}  T\textsuperscript{-1} is also  close to that in 2$H$-NbSe\textsubscript{2} ($\sim$ 0.12 $\mu$V K\textsuperscript{-1}  T\textsuperscript{-1}) \cite{Bel2003}. The same physics is very likely happening  here for CsV\textsubscript{3}Sb\textsubscript{5}. 

As shown by Wang \textit{et al.} \cite{Wangyayu2001}, the Nernst signal can be expressed as:
\begin{equation}
\begin{aligned}
S_{yx}=\frac{E_{y}}{\nabla_{x}T}&=S_{xx}\left(\frac{\alpha_{xy}}{\alpha_{xx}}-\frac{\sigma_{xy}}{\sigma_{xx}}\right)\\&=S_{xx}\left(\tan\theta_{\alpha}-\tan\theta_{H}\right)\\&=\left.\frac{\pi^{2}}{3}\frac{k_{B}^{2}T}{e}\left(\frac{\partial\,\tan\theta_{H}}{\partial E}\right)\right|_{E_{F}},
\label{eq4}
\end{aligned}
\end{equation}assuming negligible transverse thermal gradient $\nabla_{y}T$. Here, $\alpha_{ij}$ ($i,j=x,y$) and $\sigma_{ij}$ are components of the thermoelectric (Peltier)  conductivity $\overline{\alpha}$ and electric conductivity $\overline{\sigma}$ tensors, respectively. The Hall angle is defined as $\theta_H=\frac{\sigma_{xy}}{\sigma_{xx}}$ and the angle $\theta_{\alpha}=\frac{\alpha_{xy}}{\alpha_{xx}}$. For a one-band metal, if the conductivity $\overline{\sigma}$ is energy independent, i.e., $\left.\left(\frac{\partial\,\tan\theta_{H}}{\partial E}\right)\right|_{E_{F}}$=0, the two angles $\theta_H$ and $\theta_{\alpha}$ cancel out with each other (Sondheimer cancellation), leading to varnishing Nernst signal \cite{Wangyayu2001}, as found in typical metals \cite{Behnia_2009}. Such a cancellation can be avoided in a multi-band system with different types of carriers. For simplicity, we consider a two-band system consisting of electron-like and hole-like bands. The Nernst signal in Eq.(\ref{eq4}) now reads \cite{Bel2003,Behnia_2009}:
\begin{equation}
\begin{aligned}
S_{yx}=S_{xx}\left(\frac{\alpha_{xy}^{+}+\alpha_{xy}^{-}}{\alpha_{xx}^{+}+\alpha_{xx}^{-}}-\frac{\sigma_{xy}^{+}+\sigma_{xy}^{-}}{\sigma_{xx}^{+}+\sigma_{xx}^{-}}\right),
\label{eq5}
\end{aligned}
\end{equation}
with the $+$ and $-$ superscripts representing  hole-like and electron-like pockets, respectively. In general, the Sondheimer cancellation for each band does not lead to zero value in the net Nernst signal shown in Eq. (\ref{eq5}), since the signs of $\alpha_{ij}$ and $\sigma_{ij}$ now depend on the type of carriers \cite{Bel2003,Behnia_2009}. In the extreme case of compensated electron-like and hole-like bands,  $\sigma_{xy}^+=-\sigma_{xy}^-$, whereas $\alpha_{xy}^+$ and $\alpha_{xy}^-$ have the same sign and contribute additively to the Nernst effect, resulting in an enhanced $S_{yx}$ as seen in Eq. (\ref{eq5}). Note that $\alpha_{xx}^+$ and $\alpha_{xx}^-$ also have opposite signs, but do not cancel out generally.  Such kind of enhanced Nernst effect is typically observed in compensated semiconductors, which is called the ambipolar Nernst effect \cite{Delves_1965}. Examples of ambipolar Nernst effect are rarely found in metals. The  CsV\textsubscript{3}Sb\textsubscript{5} Kagome material may represent another prime metallic system to study the ambipolar effects, in addition to 2$H$-NbSe\textsubscript{2} \cite{Bel2003}.

Finally, we examine the sublinear Nernst effect appearing inside the CDW phase.  The Fig. \ref{fig3}(e) shows an enlarged view of the Nernst signal $S_{yx}$ presented in Fig. \ref{fig3}(d). Unlike the Hall data, signatures of anomalous Nernst effect are readily seen in the raw data below 30 K [see solid lines in Fig. \ref{fig3}(e) ]. This again implies that the thermoelelctric probing is very sensitive to the unusual electronic properties.  To resolve the ANE signal at higher temperatures, a local linear background has to be removed  [see dashed lines in Fig. \ref{fig3}(e) ]. Clear ANE signal appears below 40 K,  reaching maximum around 30 K, and fade away below 5 K. The non-monotonic temperature dependence of ANE rules out the magnetic origin, as the magnetization diverges at low-temperature [see Fig. \ref{fig1}(a)] in all \textit{A}V\textsubscript{3}Sb\textsubscript{5}, possibly coming from impurities\cite{Ortiz2019}. In Fig. \ref{fig3}(f), the raw data of $\alpha_{yx}=\left(\rho_{xx}S_{yx}-\rho_{yx}S_{xx}\right)/\left(\rho_{xx}^{2}+\rho_{yx}^{2}\right)$ \cite{Lixiangkang2017,YangHaiyang2020} without subtracting any background, is displayed. Here, the isotropic diagonal components $\rho_{xx}=\rho_{yy}$ and $S_{xx}=S_{yy}$  are assumed. Above 50 K, $\alpha_{yx}$ is practically linear in $B$, which becomes non-linear at lower temperatures. Clear step-like feature, together with a plateau between 1 to 2 T, are seen from 10 - 20 K, providing clear evidence of ANE effect. Notably, the maximum value of $\alpha_{yx}$ reaches $\sim$ 2.5 A m\textsuperscript{-1} K\textsuperscript{-1}, which amounts to similar magnitude of that in Co\textsubscript{3}Sn\textsubscript{2}S\textsubscript{2} \cite{YangHaiyang2020}, Co\textsubscript{2}MnGa \cite{Guin2019}, despite orders of magnitude smaller magnetization. In these magnetic topological systems, the large ANE signal was attributed to intrinsic large Berry curvature near $E_F$ \cite{YangHaiyang2020,Guin2019}.  Unlike AHE which is relevant to all occupied bands, the ANE is only sensitive to the Berry curvature around $E_F$ \cite{Xiao2006prl,XiaoRevMod}. This might explain the more evident appearance of ANE compared with AHE in  CsV\textsubscript{3}Sb\textsubscript{5}. These observations suggest that nontrivial band topology, such as the Dirac bands near the $M$ point, likely plays an important role in ANE. The emergence of ANE below 40 K imply that the multi-band nature, particularly the compensated bands, may also contribute to the observed unusual transport behaviors.

\section{Conclusions}
In summary, we have investigated magneto-Seebeck and Nernst effect in CsV\textsubscript{3}Sb\textsubscript{5}. Different charge dynamics with different energy scales above and below the CDW transition were found. The transport in the CDW phase is mainly controlled by small pockets at the $M$ point, with characteristic energy of $\sim$50 meV, leading to sizable magneto-Seebeck and magnetoresistance. The ambipolar flow of different carriers produces large Nernst signal, which peaks at the sign change temperature of the Hall coefficient, suggesting the existence of nearly compensated bands. Clear signatures of ANE emerge inside the CDW phase, which point to the important roles played by the multi-band nature and nontrivial band topology of CsV\textsubscript{3}Sb\textsubscript{5}.
\section*{Acknowledgments}

We thank Guiwen Wang at the Analytical and Testing Center of
Chongqing University for technical support. This work has been supported
by National Natural Science Foundation of China (Grant Nos.11904040,
12047564), Chongqing Research Program of Basic Research and Frontier
Technology, China (Grant No. cstc2020jcyj-msxmX0263), Fundamental
Research Funds for the Central Universities, China(2020CDJQY-A056,
2020CDJ-LHZZ-010, 2020CDJQY-Z006), Projects of President Foundation
of Chongqing University, China(2019CDXZWL002). Y. Chai acknowledges
the support by National Natural Science Foundation of China (Grant
No. 11674384, 11974065). A. Wang acknowledges the support by National
Natural Science Foundation of China (Grant No. 12004056).

\bibliographystyle{apsrev4-1}

\end{document}